\documentclass[twocolumn,showpacs]{revtex4}
\usepackage{graphicx}

\begin{document}

\preprint{APS/123-QED}

\title{Resonant Diffusive Radiation in Random Multilayered Systems}
\author{ Zh.S. Gevorkian$^{1,2}$, J.Verhoeven$^{3}$}
\address{$^{1}$Yerevan Physics Institute,
Alikhanian Brothers St. 2, Yerevan 375036, Armenia.\\
$^{2}$ Institute of Radiophysics and Electronics,Ashtarak-2,378410,Armenia.\\
$^{3}$ FOM-Instituut voor Atoom-en Molecuulfysica,Kruislaan
407,1098SJ Amsterdam, The Netherlands }

\begin{abstract}
We have theoretically shown that the yield of diffuse radiation
generated by relativistic electrons passing random multilayered
systems can be increased when a resonant condition is met.
Resonant condition can be satisfied  for the  wavelength region
representing visible light as well as soft  X-rays. The intensity
of diffusive soft X-rays for specific multilayered systems
consisting of two components is compared with the intensity of
Cherenkov radiation. For radiation at photon energy of $99.4eV$,
the intensity of Resonant Diffusive Radiation (RDR) generated by
$5MeV$ electrons passing a $Be/Si$ multilayer  exceeds the
intensity of Cherenkov radiation by a factor of $\approx 60$ for
electrons with the same energy passing a $Si$ foil. For a photon
energy of $453eV$ and $13MeV$ electrons passing $Be/Ti$ multilayer
generate RDR exceeding Cherenkov radiation generated by electrons
passing a $Ti$ foils by a factor $\approx 130$.

\end{abstract}


  \maketitle

\section{Introduction}

\indent Recently radiation generated by relativistic electrons
passing  through periodic multilayer structures have received
increased attention \cite{KLS}-\cite{LTV}.This has particularly
been enabled by the availability of multilayers with periodicities
in the nm scale. Yamada and Hosokava \cite{YH} demonstrated
Resonant Transition Radiation {RTR} \cite{Termik}-\cite{CHM} to be
applicable as a radiation source in the wavelength region
$\lambda<1nm$. This was achieved by $15MeV$ electrons passing a
periodic multilayer with $176nm$ thick nickel layers as radiator
and $221nm$ thick carbon layers as a spacer. As the RTR intensity
is proportional to $[\varepsilon_1(\omega)-
\varepsilon_2(\omega)]^2$, where $\varepsilon_1$ and
$\varepsilon_2$ are the dielectric constants of the multilayer
components, a sharp increase can be expected around inner-shell
absorption edges \cite{KLS}.

In the X-ray region the dielectric constant can be described by the
plasma formula and
usually is less than unity. However, in narrow region around
absorption edges of
some materials the dielectric constant can exceed unity. This means that the
condition for Cherenkov radiation (CR) can be fulfilled even in the
X-ray region.
The intensity of CR is proportional to $\varepsilon(\omega)-1$.

Bazylev and et al \cite{Baz} demonstrated the generation of CR
around the  $K$ absorption edge of $C$ from $1.2GeV$ electrons
passing a carbon containing foil. More recently Knulst et al
\cite{Knul} used $5MeV$ electrons to generate CR around the $Si$
$L$ absorption edge $\approx 99MeV$ and $10MeV$ electrons to
generate CR around $L$ edges of $V(512eV)$ and $Ti(453eV)$.

Previously we showed that  Diffusive Radiation(DR) generated by
relativistic electrons passing a randomly distributed multilayer
system can provide an alternative  for a source with a high
radiation yield \cite{ZHG}. Experimental evidence in favor of this
mechanism was provided by L.C.Yuan and et al \cite{Yuan} who
conducted experiments on radiation generated by relativistic
electrons passing a system of microspheres, distributed randomly
in a dielectric material. The observed strong dependence of the
radiation intensity as a function of the particle energy could
only be explained by the DR mechanism \cite{ZhCC}. In this paper we
discuss the influence of a resonance effect on the yield of
Diffusive Radiation (RDR).  As the resonant condition requires
that n exceeds 1, this effect can for radiation in the soft x-ray
region only be achieved around selected absorption edges
representing an atomic inner-shell of an element.  Our paper
explicitly deals with the generation of Resonant Diffuse Radiation
around these inner-shell absorption edges. A complicating factor
for this short wavelength region is that DR requires a low
absorption to enable multiple scattering. Nevertheless we have
shown that when a resonance condition is met related to the
anomalous behavior of dielectric constant around an adsorption
edge, the yield resulting from RDR exceeds the yield of CR in this
soft x-ray region.

\section{Initial Relations}

Let us consider  radiation from a charged particle uniformly
moving in a homogeneous medium with randomly embedded parallel
foils in it. The origin of the radiation can be explained as
follows. Each charged particle creates an electromagnetic field
around it which is not yet photon but a pseudophoton. These
pseudophotons are scattered on the inhomogeneities of dielectric
constant and convert into real photons.  We consider the radiation
due to scattering of pseudophotons which includes also the
conventional transition radiation(TR).  We do not consider the
bremsstrahlung and Cherenkov radiation which have different
origin. We have shown \cite{ZHG} that the radiation intensity can
be represented as a sum of two contributions. One is the
contribution by single scattering and the another is the
contribution caused by multiple scattering of pseudophotons. The
single scattering contribution actually is TR from randomly spaced
interfaces. However here we are interested in the contribution by
multiple scattering \cite{ZHG}, as it was shown that it leads to
the diffusion of pseudophotons. As demonstrated in \cite{ZHG} a
simplified expression for the yield for DR can be achieved by
introducing a random multilayer system with Gaussian distribution
of distances between the foils. It should be emphasised that other
random systems are not excluded to be applicable. The statistical
properties of the
Gaussian distribution enters to the solution through the elastic
mean free path of the photons.

That means that in a real system the randomness created by the
parallel foils should fulfill the requirement to treat the
dielectric constant as Gaussian distributed random function.
  For a large number of foils $N\gg 1$, in the wavelength range
$\lambda \ll l$ (where $l$ is the photon mean free path in the
normal to the foils direction, for more details see \cite{ZHG})
diffusion contribution to the radiation intensity is the main one
and given by the formula
\begin{equation}
I(\omega,\theta)=\frac{5e^2\gamma_m^2(\omega) L_z L^2 sin^2\theta}
{2\varepsilon(\omega) c l^3 |cos\theta|}
\label{inten}
\end{equation}
where $\varepsilon$ is the average dielectric constant of the system
\begin{equation}
\varepsilon=\varepsilon_0+na(\varepsilon_f-\varepsilon_0)
\label{aver}
\end{equation}
Here $\varepsilon_0$ and $\varepsilon_f$ are dielectric constants
of the medium and the plate, respectively, $a$ is the thickness of
the foils, $\gamma_m=(1-v^2\varepsilon/c^2)^{-1/2}$ is the
Lorentz factor of the particle in the medium, $L_z$ is the system
size in the $z$ direction (we assume that the particle is moving
in this direction) and $L$ is the characteristic size of the
system. The formula (\ref{inten}) has a clear physical meaning
\cite{ZHG2}. The quantity $e^2\gamma_m^2 L_z/c$ is the total
number of pseudophotons in the medium, $1/l$ is the probability of
the photon scattering and $L^2/l^2$ is the average number of
pseudophoton scatterings in the medium.
  The Eq.(\ref{inten}) is correct provided that
$\gamma_m^2\gg ak$, where $k=\omega/c$ is the photon wave number.
Taking into account that $\gamma_m^2/k$ is the coherence length
(radiation formation zone) \cite{ZHG2} this condition means that
many foils can be placed in a coherence length. Note that at the
resonance point where $\gamma_m\to \infty$, the condition
$\gamma_m^2\gg ak$ is automatically fulfilled . This regime
differs from those one which is usually explored in the periodical
multilayered systems \cite{Termik,Rul} where an enhancement of
photon yield appears as a result of constructive interference.

  Note that Eq.(\ref{inten})
is correct provided that $|cos\theta|\gg (\lambda/l)^{1/3}$. This
condition is caused by the fact that parallel to foils
pseudophotons are not diffusing so why the theory is not
applicable for very large angles. For the derivation of
Eq.(\ref{inten}) we have neglected the absorption of the
electromagnetic field. However a weak absorption $l_{in}\gg l$ (where $l_{in}$
is the photon absorption length ) can
be taken into account in the following way \cite{And}and also
\cite{ZhCC}. When $L>\sqrt{ll_{in}}$ the length $\sqrt{ll_{in}}$
becomes an effective size of the system and one should substitute
$L^2$ by $ll_{in}$ in Eq.(\ref{inten}), to obtain
\begin{equation}
I(\omega,\theta)=\frac{5e^2\gamma_m^2(\omega) L_z l_{in}(\omega) sin^2\theta}
{2\varepsilon(\omega) c l^2 |cos\theta|}
\label{abs}
\end{equation}
Note that absorption introduces an extra frequency dependence in
radiation intensity. In
section IV we will investigate the photon mean free paths in the
medium more detail.

\section{Resonant Emission}

As is seen from Eqs.(\ref{inten}) and (\ref{abs}) the intensity of
Diffusive Radiation
is proportional to the square of particle Lorentz factor in the
medium $\gamma_m^2$
\begin{equation}
\gamma_m^{-2}=1-\frac{v^2\varepsilon}{c^2}=1-\frac{v^2}{c^2}[\varepsilon_0+
(\varepsilon_f-\varepsilon_0)na] \label{lorentz}
\end{equation}
It follows from Eqs.(\ref{inten}), (\ref{abs}) and (\ref{lorentz})
that when $v$ is such that $\gamma_m^{-2}=0$ (resonant condition)
the radiation intensity drastically increases. When
$\gamma_m\to\infty$ the radiation intensity also becomes infinite.
However physically this is impossible. This infinity arises
because in theoretical consideration we assumed the system size
infinite. However in reality the system size is finite and the
intensity even at the resonance point also will be finite.
  Now consider the resonance condition $\gamma_m^{-2}=0$ in the
optical and X-ray regions,
  respectively. In the optical region
taking vacuum as a homogeneous medium, $\varepsilon_0=1$ and assuming
for relativistic
electrons that $v\approx c$,  from Eq.(\ref{lorentz}), one gets
\begin{equation}
\gamma_m^{-2}=\gamma^{-2}-\delta
\label{optic}
\end{equation}
where $\gamma=(1-v^2/c^2)^{-1/2}$ is the Lorentz factor of the
particle and $\delta=na(\varepsilon_f-1)$. So the resonance
condition in this region has the form $\gamma^{-2}= \delta$ and
weakly depends on the frequency.

Now consider the X-ray region. Assuming that
$\varepsilon_0=1-\delta_s$ and $\varepsilon_f=1-\delta_f$, for
relativistic electrons $v\approx c$, one has from
Eq.(\ref{lorentz})
\begin{equation}
\gamma_m^{-2}= \gamma^{-2}+\delta_s(1-na)+\delta_f na
\label{res}
\end{equation}
In the X-ray region one usually has $\delta_s,\delta_f>0$ and the
resonance condition $\gamma_m^{-2}=0$ can not be fulfilled.
However in the atomic inner-shell region the dielectric constant can
resonantly increase and exceed unity, \cite{KLS},\cite{Gul}. This
means that for those frequencies, for example, one can have
$\delta_f(\omega)<0$.
  As is seen from Eq.(\ref{res}) if $\delta_f(\omega)<0$ then a
resonant $\gamma_m^2\to \infty$ is possible in the X-ray region
too. The corresponding X-ray spectrum will be quite narrow because
the region where $\varepsilon_f(\omega)$ exceeds unity is small.

It should be emphasized that the resonance condition $\gamma_m^{-2}=0$ is the
Cherenkov condition for the average dielectric constant. Hence one
can say that RDR origins from the interaction of two processes.
The Cherenkov condition creates resonantly large number of
pseudophotons and multiple scattering converts them into real
photons.

Although the resonant condition is the same as for Cherenkov
radiation,the essence of RDR completely different. While
CR is based on interference of wave fronts and inhomogeneity
is not needed for its generation, DR is caused by multiple scattering of
pseudophotons on inhomogeneities. We show below that in some cases
the photon yield in RDR can exceed that one in CR.

\section{Elastic and Inelastic Mean Free Paths}

The elastic mean free paths appearing in Eqs.(\ref{inten}) and
(\ref{abs}) correspond to the  photons falling in under the normal
of the foils although those expressions determine the radiation
intensity for all angles except for a  small region . The photon
mean free path in the medium is related to the transmission
coefficient $t(\omega)$ through a foil \cite{Kos}
\begin{equation}
l(\omega)=\frac{[1-Re t(\omega)]^{-1}}{n}
\label{mfp}
\end{equation}
where $n$ is the concentration of the foils and $t(\omega)$ is given
by the formula
\begin{equation}
t(\omega)=\frac{2i
\sqrt{\frac{\varepsilon_f(\omega)}{\varepsilon_0}}\exp(-ika)}
{[\frac{\varepsilon_f(\omega)}{\varepsilon_0}+1]\sin\sqrt{\frac{\varepsilon_f(\omega)}{\varepsilon_0}}ka+
2i\sqrt{\frac{\varepsilon_f(\omega)}{\varepsilon_0}}\cos\sqrt{\frac{\varepsilon_f(\omega)}{\varepsilon_0}}ka}
\label{trans}
\end{equation}
In the Born approximation it holds that
$|\sqrt{\varepsilon_f/\varepsilon_0}-1|ka\ll 1$ even though
$(ka\gg 1)$. So one obtains from Eqs.(\ref{mfp}) and (\ref{trans})
\begin{equation}
l(\omega)=\frac{2}{n(\sqrt{\frac{\varepsilon_f(\omega)}{\varepsilon_0}}-1)^2
k^2 a^2} \label{Born}
\end{equation}
More interesting for us is the geometrical optics region
$|\sqrt{\varepsilon_f/\varepsilon_0}-1|ka\gg 1$. Substituting
Eq.(\ref{trans}) into Eq.(\ref{mfp}) and neglecting strongly
oscillating terms in the geometrical optics region, one has
$l(\omega)\approx 1/n$. Thus in this region the photon elastic
mean free path does not depend on frequency and the radiation
intensity is maximal. Usually, in the X-region
$|\sqrt{\varepsilon_f/\varepsilon_0}-1|\ll 1$ it is therefore
difficult to satisfy the condition of applicability of geometrical
optics. However, for the  atomic inner shell frequencies the
dielectric constant some times, even in the X-ray region, can
exceed  unity \cite{Gul} and for those frequencies the geometrical
optics condition $|\sqrt{\varepsilon_f/\varepsilon_0}-1|ka\gg 1$
is fulfilled. Now about the photon inelastic mean free path. In
the frequency region we are mainly interested in the dominant
mechanism of inelastic interaction of photon with the medium is
the photoabsorption. Therefore the inelastic mean free path can be
represented in the form
\begin{equation}
l_{in}^{-1}=N_1\sigma_{1a}(\omega)+ N_2\sigma_{2a}(\omega)
\label{inel}
\end{equation}
where $N_{1,2},\sigma_{1,2}$ are the numbers of atoms in an unit
volume and photoabsorption cross section for the first and second
media, respectively.

\section{Absorption of RDR photons in the Medium}

Above we have discussed the absorption of pseudophotons in the
random medium. However already formed real photons also will be
absorbed in the medium. Therefore to know what part of already
created real photons will escape the system one should take into
account the absorption of real photons in the medium. Suppose that
we are interested in the photon yield from the depth $z$ in the
material. Using Eq.(\ref{abs}) and adding an exponential decaying
factor which takes into account the difference of paths of photons
with different emitted angles, one has
\begin{equation}
\frac{dI}{dz}=\frac{5e^2\gamma_m^2(\omega)  l_{in}(\omega)
sin^2\theta} {2\varepsilon(\omega) c l^2
|cos\theta|}\exp\left(-\frac{z}{l_{in}|cos\theta|}\right)
\label{absrp}
\end{equation}
Here $\frac{dI}{dz}$ is the spectral-angular radiation intensity
per unit length of electron path in the medium. Note the
suppression of radiation intensity at very large angles. The real
DR photons are formed in the effective size $(ll_{in})^{1/2}$.
Therefore when finding the total radiation intensity one should
cut the integral in the lower limit on this length.After
integration for total spectral angular intensity, we have
\begin{equation}
I=\frac{5e^2\gamma_m^2(\omega)  sin^2\theta l_{in}^2}
{2\varepsilon(\omega)
cl^2}\exp\left[-\left(\frac{l}{l_{in}}\right)^{1/2}\frac{1}{|cos\theta|}\right]
\label{absrpt}
\end{equation}
So as one could anticipate the absorption of real DR photons leads
to the cutoff of radiation intensity at large angles and maximum
lies at medium angles.

\section{Qualitative Estimates for Specific Elements}

We have shown that  resonant emission of radiation intensity is
possible in a system with one-dimensional randomness of the
dielectric constant. Resonant emission is possible in the optical
as well as in the X-ray regions. The latter case is most
interesting because it forms high bright  soft X-ray source using MeV
electrons, \cite{KLS}, \cite{LTV}.

One of the main conditions for realization of diffusive radiation
mechanism is $l_{in}\gg l$. The minimal value of $l$ is reached in
the geometrical optics region, $l\sim 1/n$. The geometrical optics
approximation for $l$ is justified provided that
$|\sqrt{\varepsilon_f/\varepsilon_0}-1|ka\geq 1$. In order to
obtain an impression how large the yield of DR photons can be let
us make a crude comparison  with  Cherenkov radiation. As is shown
in \cite{ZHG2} the ratio of DR and Cherenkov radiation intensities
under weak absorption is of order
\begin{equation}
\frac{I^D}{I^{C}}\sim \frac{1}{kl}\gamma_m^2 \frac{l_{in}}{l}
\label{compar}
\end{equation}
It should be emphasized that this formula represents a leading
order comparison for the yield RDR and CR. When the resonance
condition $v \to c/\sqrt{\varepsilon}$ is fulfilled, CR as well as
RDR within the same wavelength are generated. While the emission
angle of Cherenkov radiation is fixed at
$cos\theta=c/v\sqrt{\varepsilon}$, RDR is emitted within a wide
angular distribution. We omit the angular part in (\ref{compar})
for simplicity.
  Consider the conditions of generation of RDR
for specific structures using the data from \cite{site}. First
consider the structure  $Be/Si$ at the photon energy
$\omega=99.4eV$ near the L3 edge of $Si$. Note that we take the
photon energy at the long wavelength side of  the $Si$  $L3$
$\omega=99.8eV$ edge where the absorption weaker. However for
$99.4eV$ the refractive index $n_{Si}(\omega)=1+11.56\times
10^{-3}$ still exceeds unity. Using also that $n_{Be}=1-4.27\times
10^{-3}$  one can be convinced that the geometrical optics
approximation is valid provided that $a>130nm$. The photon
attenuation length at $\omega=99.4eV$ in  $Si$ and $Be$ is
$0.55\mu m$ and $0.76\mu m$, respectively \cite{site}. Therefore
if one takes the average distance $d$ between $Si$ foils $\sim 130
nm$, the elastic and inelastic mean free paths of photon
approximately can be estimated as $l\sim d\sim 130nm$ and
$l_{in}\sim 0.62\mu m$. As a result the photon multiple scattering
condition $l_{in}\gg l$ is fulfilled. Now let us find the resonant
electron energy. Using the above mentioned values of refraction
index for  $Be$ and $Si$, assuming the filling factor of  $Si$
approximately, $na\sim 1/2$, one finds from Eq.(\ref{res}),that
$\gamma_r=11.71$. Correspondingly the resonant electron energy
will be $E_r=5.355MeV$. Now estimate the ratio $I^D/I^{C}$ using
Eq.(\ref{compar}). For a photon energy $\omega=99.4eV$ for
$l=130nm$, the factor $1/kl\approx 0.01$. As at the resonance
point our formulae are not applicable we choose an electron energy
of $E=5.1MeV$ and find  $\gamma_m^2\approx 1471$. Finally, taking
into account that $l_{in}/l\sim 4$, one finds a much larger
diffusive radiation than Cherenkov radiation $I^D/I^{C}\sim 60$.
The number of foils and correspondingly the system size  can not
be too large because of absorption. On the other hands one should
have enough number of foils to treat the system as a random. For
the average photon attenuation length $0.6\mu m$ a system size of
order $10\mu m$ is reasonable. This size means approximately $40$
foils. Another interesting possibility is to generate  RDR around
the $L3$ edge of $Ti$. For a photon energy of $\omega=453eV$ the
refraction index of $Ti$ is \cite{site} $n_{Ti}=1+3.0826\times
10^{-3}$. As a spacer material choose a light element to make
reduce the total absorption. Let consider, for example, $Be$. For
$Be$ at $453eV$ $n_{Be}=1-1.7932\times 10^{-3}$ and attenuation
length is $0.9117\mu m$. The attenuation length for $Ti$ at
$453eV$ is $0.661\mu m$. Analogous calculations as for the $Be/Si$
combination show that the geometrical optics condition
$ka|n_{Ti}/n_{Be}-1|\geq 1$ can be satisfied for $100nm$ thick
$Ti$ foils. The resonance electron energy for a $Ti$ filling
factor $na\sim 1/2$ is $13.45MeV$. Taking for the electron energy
$E=13MeV$ one obtains for the ratio $I^D/I^C\sim 130$. The number
of foils could be $40-50$ and the system size $10\mu m$.

\section{Discussion}
Let us first point out the main differences between random and periodical
systems.
Gaussian averaging over the randomness enables one to obtain a compact
expression for radiation yield in contrary to the sum of complicated
terms in periodical case. It is easier to construct a random system than a
periodical one therefore random systems are more convenient from practical
point of view. Consideration of pseudophoton multiple scattering effects
in periodical systems remains as an open problem.

Many approximations are used when we estimate the ratio $I^D/I^C$.
Therefore we emphasize that the calculated values for $I^D/I^C$ do
not represent exact optimized values.
The calculations only give an impression about a possible yield.
The before mentioned ratios are very sensitive to electron energy,
refractive index, filling factor $na$, frequency etc. because the
resonant character of radiation. Theoretically for the resonance
condition the ratio can be very large. However this requires
electrons in the MeV range with an energy accuracy of better than
$0.01MeV$. As shown before, RDR for the Extreme Ultra Violet and
shorter wavelength regions is only possible within a narrow
bandwidth for which $n(\omega)>1$.

\section{Conclusions}

The possibility to fundamentally generate Resonant Diffusive
Radiation for the short wavelength region (EUV and soft x-rays)
using a multilayer system with random periodicity has been
demonstrated theoretically. For a resonant condition it is
required that $\gamma_m^{-2}=0$. For the wavelength region as
considered, this is only possible around absorption edges of some
elements where $\varepsilon>1$.

We demonstrated the feasibility to apply this phenomenon on a
practical bright source. Therefore we made some preliminary
calculations for specific materials combinations, comparing the
RDR yield with the CR yield. Experimental research is justified to
further understanding of the phenomenon. Moreover, we expect the new
high brightness photon sources can be
developed, applicable in x-ray microscopy or EUV lithography.

\section{Acknowledgements}
We are indebted to M.J.van der Wiel for helpful discussions.
  This work of Zh.S.G., was supported by a grant
A-655 from International Science and Technology Center. This work
of J.V. is carried out as part of the research programme of the
Stichting voor Fundamenteel Onderzoek der Materie (FOM) with
financial support from the Nederlandse Organisatie voor
Wetenschappelijk Onderzoek (NWO).

\end{document}